\documentclass[pra,aps,twocolumn,superscriptaddress,showpacs]{revtex4}
\usepackage{epsfig}
\usepackage{amsmath}

\begin{document}

\title{Role of  Particle Interactions in the  Feshbach Conversion of
Fermion Atoms  to Bosonic Molecules }
\author{Jie Liu}
\affiliation{Institute of Applied Physics and Computational
Mathematics, Beijing, 100088, China}\affiliation{Center for Applied
Physics and Technology, Peking University, 100084, Beijing,
P.R.China}
\author{Li-Bin Fu}
\affiliation{Institute of Applied Physics and Computational Mathematics, Beijing, 100088,
China}
\author{Bin Liu}
\affiliation{Institute of Applied Physics and Computational
Mathematics, Beijing, 100088, China}
\author{Biao Wu}
\affiliation{Institute of Physics, Chinese Academy of Sciences, P.O.Box 603, Beijing
100080, China}

\begin{abstract}
We investigate the Feshbach conversion of fermion atomic pairs
to condensed boson molecules with a microscopic model that accounts
the repulsive interactions among all the particles involved. We find
that the conversion efficiency is enhanced by the interaction between
boson molecules while suppressed by the interactions between fermion
atoms and between atom and molecule. In certain cases, the combined
effect of these interactions leads to a ceiling of less than $100\%$
on the conversion efficiency even in the adiabatic limit. Our model
predicts a non-monotonic dependence of the efficiency on mean atomic
density. Our theory agrees well with recent experiments
on $^6$Li and $^{40}$K.
\end{abstract}
\date{\today}
\pacs{03.75.Ss, 05.30.Fk, 05.30.Jp, 03.75.Mn}
\maketitle
Feshbach resonance has now become a focal point of the research
activities in cold atom physics\cite{Timmermans,Stoof,chen,Julienne}
since its first experimental realization\cite{Inouye}.
Among these research activities, the production of
diatomic molecules from Fermi atoms with Feshbach resonance is of
special interest and has attracted great attention. First, it is an
interesting phenomenon by itself; second, it provides a unique
experimental access to the BCS-BEC crossover physics\cite{jin}.
So far, by slowly sweeping the magnetic field through the Feshbach
resonance, samples of over $10^5$ weakly bound molecules
(binding energy $\sim 10$ kHz) at temperatures of a few tens of nK
have been produced from quantum degenerate
Fermi gas\cite{regal,strecker,cubizolles}.

The Feshbach conversion is a complicated process involving many
fermion atoms and boson molecules in a sweeping magnetic field that
crosses a resonance. The theoretical description of the conversion
efficiency as a function of sweep rate, atom mass, atomic density,
and temperature is still under development. The existing theories
include  the Landau-Zener (LZ) model of two-body molecular
production\cite{lz,goral} and its  many-body extension at zero
temperature\cite{vardi1,vardi2,altman}, phase-space density
model\cite{hodby}, and equilibration model at finite
temperatures\cite{williams} .

In this Letter we study a microscopic model of the Feshbach conversion
that accounts all the two-body interactions, which include atom-atom,
molecule-molecule, and atom-molecule interactions. These interactions
are ignored in previous theoretical studies\cite{goral,vardi1,vardi2,altman}.
We find that these interactions affect strongly the Feshbach
conversion efficiency: the repulsive interaction between molecules
tends to enhance the conversion efficiency while the other two
repulsive interactions between atoms and between atom and molecule
suppress the efficiency. Combined together, these interactions
can yield a ceiling of less than $100\%$ for the conversion efficiency
even in the adiabatic limit of the sweeping magnetic field.
This interaction-suppressed conversion efficiency is in spirit
the same as the broken adiabaticity by interaction in the nonlinear
LZ tunneling\cite{wu,nlz}. In addition, our model predicts a
non-monotonic dependence of the conversion efficiency on mean
atomic density. Our results are compared to recent experiments
with $^6$Li and $^{40}$K\cite{strecker,hodby}; they are
in good agreement.

To include all particle interactions, we extend the two channel
model\cite{model1,model2,model3} and write the Hamiltonian as
\begin{eqnarray}
H &=&\sum_{\mathbf{k},\sigma }\epsilon _{\mathbf{k}}a_{\mathbf{k},\sigma
}^{\dagger }a_{\mathbf{k},\sigma }+\left( \gamma +\frac{\epsilon _{b}}{2}
\right) b^{\dagger }b
\notag \\
&&+\frac{ U_a}{V_a} \sum_{\mathbf{k},\mathbf{k^{\prime
}}}a_{\mathbf{k},\uparrow }^{\dag }a_{-\mathbf{k},\downarrow }^{\dag
}a_{-\mathbf{k^{\prime }},\downarrow }a_{\mathbf{k^{\prime }}
,\uparrow }    \notag \\
&& +\frac{U_{ab}}{V_{a}}\sum_{\mathbf{k},\sigma }a_{\mathbf{k},\sigma
}^{\dagger }a_{\mathbf{k},\sigma }b^{\dagger }b
+\frac{U_b}{V_{b}}b^{\dagger }b^{\dagger }bb \notag \\
&& +\frac{gV_{b}}{V_{a}^{3/2}}\sum_{\mathbf{k}}\left( b^{\dagger
}a_{-\mathbf{k} ,\downarrow }a_{\mathbf{k},\uparrow
}+a_{\mathbf{k},\uparrow }^{\dagger }a_{- \mathbf{k},\downarrow
}^{\dagger }b\right)\,.
\label{ham}
\end{eqnarray}
Here $\epsilon _{\mathbf{k}}=\hbar ^{2}k^{2}/2m_{a}$ is the kinetic
energy of the atom and $\sigma=\uparrow,\downarrow$ denote the two
hyperfine states of the atom. $U_a=\Lambda U_0,U_{ab}=\Lambda
U_1,g=\Lambda g_0$, and $\gamma=\gamma_0-\Lambda g_0^2/U_c$, where
$\gamma_0 =\mu _{co}(B-B_{0})$ is the molecule energy under the
linearly changing magnetic field with $B=-\alpha _{r}t$,
$g_{0}=\sqrt{4\pi \hbar ^{2}a_{bg}\Delta B\mu _{co}/m_{a}}$ is the
atom-molecule coupling \cite{chen},  $U_{0}=4\pi \hbar
^{2}a_{bg}/m_{a}$ is the interaction between atoms, $U_{1}=4\pi
\hbar ^{2}1.2a_{bg}/m_{ab}$ is the atom-molecule scattering
interaction, and $U_{b}=4\pi \hbar ^{2}0.6a_{bg}/m_{b}$  is the
interaction between molecules\cite{scatter}. $\Lambda \equiv
(1+U_0/U_c)^{-1}$ and $U_c^{-1}
=-\sum_k(\exp(-k^2/k_c^2)/2\epsilon_k)$ with the cutoff momentum
$k_c$ representing the inverse range of interaction
\cite{model3,kokk,chen1}. $B_{0}$ and $\Delta B$ are Feshbach
resonance point and width, respectively. $m_{a}$ and $m_{b}=2m_{a}$
are masses for atoms and molecules, and
$m_{ab}=\frac{2}{3}m_{a}$ is the reduced mass for the atom-molecule
interaction.

In experiments, the molecular bosons are more tightly confined in
space than the fermion atoms due to their different statistics. To
reflect this, we use $V_a$ for  the volume of fermion atoms and
$V_b$ for bosonic molecules. We assume the zero temperature limit,
consider only one bosonic mode, and ignore all possible dissipations
in the system, such as the loss of atoms by three-body collision.

In current experiments, the intrinsic energy width of a Feshbach
resonance is much larger than the Fermi energy $E_F$\cite{Diener},
it is therefore reasonable to assume $\epsilon_{\mathbf{k}}=\epsilon$.
This is called degenerate model in Ref.\cite{model2,vardi1,vardi2}.
We introduce the following operators\cite{vardi1,vardi2},
$L_{x} = \frac{\sum_{\mathbf{k}}(a_{%
\mathbf{k},\uparrow}^{\dagger}a_{-\mathbf{k},\downarrow}
^{\dagger}b+b^{\dagger}a_{-\mathbf{k},\downarrow} a_{\mathbf{k},\uparrow})}{%
(N/2)^{3/2}} , L_{y} = \frac{\sum_{\mathbf{k}}(a_{\mathbf{k}%
,\uparrow}^{\dagger}a_{-\mathbf{k},\downarrow} ^{\dagger}b-b^{\dagger}a_{-%
\mathbf{k},\downarrow} a_{\mathbf{k},\uparrow})}{i(N/2)^{3/2}}, L_{z} =\frac{
\sum_{\mathbf{k},\sigma}a_{\mathbf{k},\sigma}^{\dagger} a_{\mathbf{k}%
,\sigma}-2 b^{\dagger}b}{N}$, where $N=2 b^{\dagger}b+\sum_{\mathbf{k}%
,\sigma}a_{\mathbf{k},\sigma}^{\dagger} a_{\mathbf{k},\sigma}$ is the
total number of atoms. The Hamiltonian becomes
$H=\frac{N }{4}\left(2\epsilon-(\gamma+\frac{\epsilon_b}{2})-
\frac{N U_a}{2 V_a}-\frac{N U_{ab}}{V_{a}}\right) L_z %
-\frac{N^2}{16}\left(\frac{U_a}{V_a}+\frac{2U_{ab}}{V_a}-
\frac{U_b}{V_b}\right)\left(1-L_z\right)^2
+\frac{gV_{b}}{V_{a}^{3/2}}\left(\frac{N}{2}%
\right)^{3/2} L_x$ \cite{reduce}.
With the commutators $[L_z, L_x]=\frac{%
4i}{N}L_y,[L_z, L_y]=-\frac{4i}{N} L_x,[L_x, L_y]=\frac{i}{N}(1-L_z)(1+3
L_z) + o(1/N^2)$, we can obtain the Heisenberg equations for the system
$i\hbar\frac{\mathit{d}}{\mathit{d}t}L_l=[L_l, H], (l=x,y,z)$. Since all the
commutators vanish in the limit of $N\rightarrow\infty$ and $N$
is large in current experiments, it is appropriate to take $L_x$, $L_y$,
and $L_z$ as three real numbers $u,v,w$, respectively. These
Heisenberg equations are then reduced to
\begin{eqnarray}
\label{cannon1}
du/d\tau&=&-\delta v-2\chi v(1-w)\,,~~~dw/d\tau=\sqrt{2}v\,,\\
\label{cannon2}
dv/d\tau&=&\frac{3\sqrt{2}}{4}(w-1)(w+\frac{1}{3})
+\delta u+2\chi u(1-w)\,,
\end{eqnarray}
where $\tau=(g V_b\sqrt{N}/\hbar V_a^{3/2})t$. Because of the
identity $u^2+v^2=\frac{1}{2}(w-1)^2(w+1)$, with introducing the
canonical variable  $\theta=\arctan(v/u)$ we have  a classical
Hamiltonian,
\begin{eqnarray}
\mathcal{H} &=& \delta w-\chi (1-w)^2
+\sqrt{(w-1)^2(w+1)}\cos\theta. \label{clas}
\end{eqnarray}
The above equations show that all the experimental parameters affect the system
via only two dimensionless parameters $\delta$ and $\chi$.
By a trivial shift of time origin, we can set $\delta=\alpha \tau$ with
\begin{equation}
\label{alpha}
\frac{\alpha_r}{\alpha} = \frac{4\pi\hbar n a_{bg} \Delta
B}{m_a}\Lambda^2\frac{V_b^2}{V_a^2}\,,
\end{equation}
where $n=N/V_a$ is the  mean atomic density.
The nonlinear parameter $\chi$ is given by
\begin{equation}
\chi =\left(2.3-\frac{0.15 V_a}{\Lambda V_b}\right)
\frac{V_a}{V_b}\sqrt{\frac{\pi\hbar^2
a_{bg}n}{m_a\mu_{co}\Delta B}}\,.
\label{chi}
\end{equation}
\begin{figure}[!ht]
\centering
\includegraphics[width=0.5\textwidth]{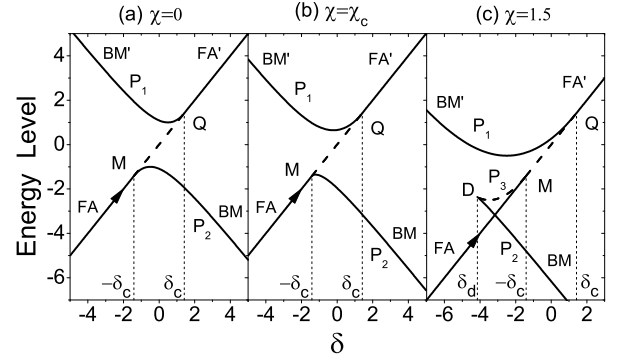}
\caption{Adiabatic energy levels for different interaction strengths. (a)$%
\protect\chi=0$; (b)$\protect\chi=\protect\chi_c=\protect\sqrt{2}/4$; (c)$%
\protect\chi=1.5$. The unstable states are indicated by dashed lines
(MQ and DM).} \label{levels}
\end{figure}

To understand the dynamics, we first look at the fixed points
$\dot{w}= \dot{u}=\dot{v}=0$ of Eqs.(\ref{cannon1},\ref{cannon2}).
The energies for these fixed points make up energy levels of the
system as shown in Fig.\ref{levels}. One sees that the structure of
these energy levels changes dramatically as the nonlinear parameter
$\chi$ increases. Specifically, we observe: (\textit{i}) There are
two fixed points when $|\delta|$ is large enough: one for bosonic
molecule (BM) and the other for fermion atom (FA). (\textit{ii})
When $|\delta|<\delta_c=\sqrt{2}$, there is an additional fixed
point with $w=1$. However, this fixed point is dynamically
unstable\cite{nlz}. (\textit{iii}) For $\chi>\chi_c=\sqrt{2}/4$,
there appears one more fixed point denoted by $P_3$ and,
consequently, a loop in the energy levels. As we shall see, this
loop has highly non-trivial physical consequences. This fixed point
$P_3$ is also unstable.

Consider the adiabatic evolution of the system starting from a high
negative  value of $\delta$ with $w=1$. This corresponds to the
experiments where the magnetic field sweeps slowly across the
Feshbach resonance with no bosonic molecules initially. When $\chi$
is small, such as in Fig.\ref{levels}(a), the evolution of the
system follows the solid line, converting all fermion atoms into
molecules. However, when $\chi$ is beyond $\chi_c$ as in
Fig.\ref{levels}(c), the system will find no stable energy level to
follow at singular point $M$. As a result, only a fraction of
fermion atoms are converted  into bosonic molecules.

\begin{figure}[!ht]
\centering
\includegraphics[width=7cm]{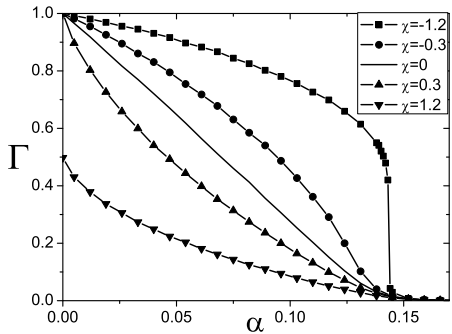}
\caption{Conversion efficiency  $\Gamma$ as a function of the
sweeping rate $\protect\alpha$ for various interactions.}
\label{power}
\end{figure}

This simple analysis is confirmed by our numerical results, which
are plotted in Fig.\ref{power}. In our calculation, the 4-5th
Runge-Kutta step-adaptive algorithm is used in solving the
differential equations (\ref{cannon1},\ref{cannon2}). Because $w=1$
is a fixed point when $\delta <-\sqrt{2}$, we start from
$(w,u,v)\approx(1,0,0)$  and sweep the field from $\delta
=-\sqrt{2}$ to $200$.  In Fig.\ref{power}, the conversion efficiency
$\Gamma$, i.e., the fraction of the converted fermion atom pairs is
drawn as a function of $\alpha$. Evidently, $\Gamma$ approaches one
as $\alpha\rightarrow 0$ when $\chi<\chi_c$, indicating that all
atomic pairs are converted into molecules. In contrast, when
$\chi>\chi_c$, $\Gamma$ does not increase to one in the adiabatic
limit $\alpha\rightarrow 0$. This means that there is a ceiling
$\Gamma_{ad}$($<100\%$) on the conversion efficiency. Moreover,
Fig.\ref{power} demonstrates that positive $\chi$ suppresses the
conversion efficiency whereas the negative $\chi$ enhances it.
Because the repulsive interaction between bosonic molecules enters
$\chi$ as a negative value, it enhances the conversion efficiency;
the repulsive fermion atomic interaction and atom-molecule
interaction contribute positively to $\chi$, they suppress the
conversion.

\begin{figure}[!t]
\centering
\includegraphics[width=0.5\textwidth]{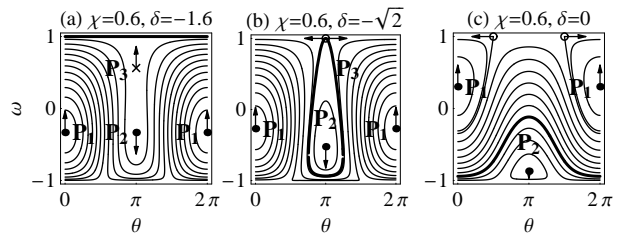}
\caption{Phase spaces of Hamiltonian (\protect\ref{clas}).
 The dark line in (a) is for the fixed
point $w=1,u=0,v=0$. It is a line because $\theta$ is not defined
at $u=v=0$. The two fixed points on line $w=1$ in (c)
are in fact the same fixed point; they are artifact caused by the definition
$\protect\theta=\arctan(v/u)$.} \label{phase}
\end{figure}

The ceiling  $\Gamma_{ad}$ on the atom-molecule
conversion efficiency depends on $\chi$. This dependence can be
found by examining the phase space diagrams of our system
shown in Fig.\ref{phase}. As $\delta$ ramps up slowly from a
large negative value, the fixed point $P_3$ will move up
until it hits the fixed point $w=1,u=0,v=0$, represented by
a dark straight line in Fig.\ref{phase}(a). This collision
occurs at $\delta=-\sqrt{2}$. Immediately after the collision,
the hyperbolic fixed point $P_3$ is no longer a fixed point and
becomes a solution that evolves along the dark
line in Fig.\ref{phase}(b). The dark line is given by $\sqrt{2}=\chi (1-w)-
\sqrt{1+w}\cos \theta$, which is found by taking $E=\delta=-\sqrt{2}$ in
the Hamiltonian (\ref{clas}). As the action of this trajectory is
nonzero while a fixed point has zero action, this collision of the
two fixed points represents a sudden jump in action. It is this sudden jump
that has caused the nonzero fraction of remnant atoms. As $\delta$ ramps
up further slowly, the trajectory will change its shape as witnessed
in Fig.\ref{phase}(c); however, its action stays constant as demanded by the classical
adiabatic theorem\cite{lan,lwn}. The action is
\begin{equation}
I =\begin{cases}
 \frac{1}{2 \pi} \oint \frac{\cos\theta \sqrt{8\chi^2-4\sqrt{2}%
\chi+\cos^2\theta}}{2\chi^2} \mathit{d}\theta ,
& \frac{\sqrt{2}}{4}<\chi<\frac{\sqrt{2}}{2};\\
 \frac{1}{2 \pi} \int_0^{2\pi}\frac{4\chi^2-2\sqrt{2}\chi+\cos^2\theta}{2\chi^2}
 \mathit{d}\theta ,
& \chi>\frac{\sqrt{2}}{2},
\end{cases}
\end{equation}
which yields the ceiling on the efficiency
\begin{equation}
\Gamma_{ad}=1-\frac{1}{2}I = \frac{ 4\sqrt{2}\chi-1}{ 8\chi^2},
\chi>\frac{\sqrt{2}}{4}; \quad \Gamma_{ad}=1, \chi
<\frac{\sqrt{2}}{4}. \label{gct}
\end{equation}

\begin{figure}[!ht]
\centering
\includegraphics[width=0.5\textwidth]{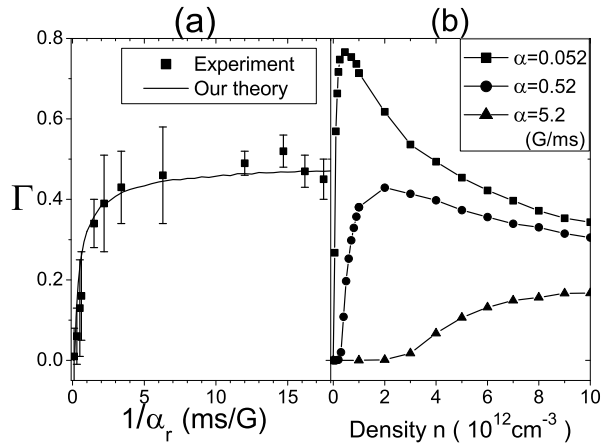}
\caption{(a)Comparison between our theory and experimental data
of $^6$Li\protect\cite{strecker} for the conversion efficiency
$\Gamma$ as a function of field sweep rates. (b)
The dependence of $\Gamma$ on the mean atomic density.}
\label{comp}
\end{figure}

Now we compare our theory with existing experiments. For the
experiment with $^{6}$Li\cite{strecker}, the mean density is
$n=4\times 10^{12}{\rm cm}^{-3}$ with $N=6\times 10^{5}$ atoms. The
scattering length $a_{bg}=59 a_{B}$, $\mu _{co}\sim 2\mu _{B}$,
where $a_{B}$ and $\mu _{B}$ are Bohr radius and Bohr magneton,
respectively, and the resonance width $\Delta B=0.1$G at
$B_{0}=543.8$G. The Fermi energy $E_F$ in the combined harmonic and
box-like trapping potential of Ref.\cite{strecker} is given by
$E_F=\left[15\pi N \hbar^3
\omega_r^2/\left(8\sqrt{2m_a}L\right)\right]^{2/5}$, where
$\omega_r=2\pi\times800 {\rm s}^{-1}$ is the angular frequency of
the radial harmonic trap and $L=480\mu$m is the size of the axial
potential. The ground state energy of molecular bosons is $E_G=\hbar
\omega_r+\hbar^2\pi^2/(2m_bL^2)$. Then we have $V_a/V_b=E_F/E_G=36$.
We set $\Lambda = 391$ with a momentum cutoff
$K_c=96k_{F}$\cite{note}. From Eq.(\ref{alpha}), the sweeping rate
is $\alpha_r/\alpha=20{\rm G/ms}$. The second term in the bracket of
Eq.(\ref{chi}) that accounts for the repulsive interaction between
bosonic molecules is small,  so the interaction parameter takes the
form of $\chi\simeq  2.3\frac{V_a}{V_b} \sqrt{\frac{\pi\hbar^2
a_{bg}n}{m_a\mu_{co}\Delta B}}$. From the above experimental
parameters, we find the interaction parameter as $\chi =1.26$. This
strong interaction ($> \chi_c$) indicates a ceiling of $\Gamma
_{ad}=0.48$ via Eq.(\ref{gct}). This is in good agreement with
experiments (see Fig.\ref{comp}a).

For $^{40}{\rm K}$, the situation is different. The resonance at
$B_0=202.1$G has a large width of $\Delta B =7.8$G and the mass of
$^{40}$K is 7 times that of $^6$Li. In Ref.\cite{hodby}, the
fermions are confined in a dipole trap characterized by a radial
frequencies $\nu_r$ between 312 and 630 Hz and an aspect ratio of
$\nu_r/\nu_z=70$. The Fermi energy is $E_F=\hbar
\left(3N\omega_r^2\omega_z\right)^{1/3}$ and the ground state energy
of condensed bosons is $E_G=\hbar\omega_r+\hbar\omega_z/2$. For the
dipole trap, the ratio  $V_a/V_b=\left(E_F/E_G\right)^{3/2}=251$.
With $a_{bg}=174 a_B$, $\mu_{co} \sim 2\mu_B$, initial clouds have
mean densities $n=2\times 10^{12}$cm$^{-3}$, and $N=2.5\times
10^5$\cite{hodby}, we obtain $\chi=0.135\sim0.274$, which is less
than the threshold $\chi_c=\sqrt{2}/4$. Therefore, $^{40}{\rm K}$
atom pairs can be completely converted to bosonic molecules in
adiabatic limit. Indeed, the conversion efficiency up to $90\%$ has
been observed\cite{hodby}.

We emphasize that the suppressed conversion efficiency by particle
interaction dominates only at low temperatures. As a result, in the
above we have only compared to the data obtained at low temperatures
($T/T_F=0.1$ for $^6$Li and $T/T_F=0.05$ for $^{40}{K}$).
Temperature can affect the conversion efficiency strongly as reported
in Ref. \cite{hodby}. The ceiling of $50\%$ conversion efficiency
observed in Ref.\cite{regal} is likely a thermal effect since the
experiment is performed at $T/T_F =0.33$, and has been  explained by the theories
of finite temperature\cite{pazy,Chwe}

For the $^6$Li,  from Eq.(2)(3)we have also calculated numerically
the conversion efficiency as a function of sweeping rate.  The
comparison between our theory and experiment is shown in
Fig.\ref{comp}a. They are in a good agreement. In addition, our
model predicts a non-monotonic dependence of the conversion rate on
the mean atomic density (see Fig.\ref{comp}(b)). This can be
understood from Eqs.(\ref{alpha},\ref{chi}). In Eq.(\ref{alpha}), we
see the effective sweeping rate $\alpha$ is inversely proportional
to the atomic density. So, increasing the density will reduce the
effective sweeping rate and therefore enhance the conversion rate.
On the other hand, higher density will give larger nonlinearity
$\chi$ as indicated in Eq.(\ref{chi}), which in turn suppresses the
atom-molecule conversion. These two factors compete with each other,
giving rise to the non-monotonic curves in Fig.\ref{comp}(b). In
practical experiments, to achieve higher conversion efficiency, one
needs to carefully choose initial fermion atom density, making it
fall into the optimal parameter regime.

In summary, we have identified the significant role of the
interactions between particles in the Feshbach conversion of atomic
fermion pairs to molecular bosons. Our theory is consistent with the
existing experiments. Our model also predicts a non-monotonic dependence
of the conversion rate on the mean atomic density, which is
important for the optimal choice of parameters in future Feshbach
experiments.

This work was supported by NSF of China (10725521,10604009,10504040)
and the 973 project (2006CB921400,2007CB814800). B.W. is also
supported by the ``BaiRen'' program of the CAS.



\begin{thebibliography}{99}
\bibitem{note} The cutoff $k_c$ is chosen such that the tunnelling
window $2\delta_c$ for converting  atomic fermions to molecular
bosons  is consistent with the Feshbach resonance width $\mu_{co}\Delta B$ of
$^6$Li.
\end{thebibliography}
\end{document}